\documentclass[journal]{IEEEtran}

\usepackage[english]{babel}
\usepackage{graphicx, grffile}% Include figure files
\usepackage[font=normal]{subfig}
\usepackage{dcolumn}% Align table columns on decimal point
\usepackage{physics}
\usepackage{bm}% bold math
\usepackage{hyperref}% add hypertext capabilities
\usepackage{cleveref}
\usepackage{fancyhdr}
\usepackage{romannum}% add roman number
\usepackage{enumerate}
\usepackage{tabularx}
\usepackage{floatrow}
\usepackage{caption}
\usepackage{setspace}
\usepackage{comment}
\usepackage{multicol}
\usepackage{flushend}
\usepackage{scalerel}
\usepackage{tikz}
\usetikzlibrary{svg.path}

\definecolor{orcidlogocol}{HTML}{A6CE39}
\tikzset{
  orcidlogo/.pic={
    \fill[orcidlogocol] svg{M256,128c0,70.7-57.3,128-128,128C57.3,256,0,198.7,0,128C0,57.3,57.3,0,128,0C198.7,0,256,57.3,256,128z};
    \fill[white] svg{M86.3,186.2H70.9V79.1h15.4v48.4V186.2z}
                 svg{M108.9,79.1h41.6c39.6,0,57,28.3,57,53.6c0,27.5-21.5,53.6-56.8,53.6h-41.8V79.1z M124.3,172.4h24.5c34.9,0,42.9-26.5,42.9-39.7c0-21.5-13.7-39.7-43.7-39.7h-23.7V172.4z}
                 svg{M88.7,56.8c0,5.5-4.5,10.1-10.1,10.1c-5.6,0-10.1-4.6-10.1-10.1c0-5.6,4.5-10.1,10.1-10.1C84.2,46.7,88.7,51.3,88.7,56.8z};
  }
}

\newcommand\orcidicon[1]{\href{https://orcid.org/#1}{\mbox{\scalerel*{
\begin{tikzpicture}[yscale=-1,transform shape]
\pic{orcidlogo};
\end{tikzpicture}
}{|}}}}
\newcommand{\myref}[1]{(\ref{#1})}

\graphicspath{{figure/}}

\begin{document}

\pagenumbering{arabic}

%\singlespacing
\title{Divide-and-conquer policy in the Naming Game}

\author{Cheng Ma \orcidicon{0000-0002-3586-7750}, Brendan Cross \orcidicon{0000-0001-5060-3705}, Gyorgy Korniss \orcidicon{0009-0005-6816-6458}, Boleslaw K. Szymanski\orcidicon{0000-0002-0307-6743}, \IEEEmembership{Life Fellow, IEEE}
\thanks{Cheng Ma, Gyorgy Korniss, and Boleslaw K. Szymanski are with the Department of Physics, Applied Physics and Astronomy, Rensselaer Polytechnic Institute, Troy, NY 12180, USA}
\thanks{Brendan Cross and Boleslaw K. Szymanski are with the Department of Computer Science, Rensselaer Polytechnic Institute, Troy, NY 12180, USA}
\thanks {All authors are also with Network Science and Technology Center, Rensselaer Polytechnic Institute, Troy, NY 12180, USA}
}

%\markboth{IEEE TRANSACTIONS ON COMPUTATIONAL SOCIAL SYSTEMS  }

\maketitle

\begin{abstract}
The Naming Game is a classic model for studying the emergence and evolution of language within a population. In this paper, we extend the traditional Naming Game model to encompass multiple committed opinions and investigate the system dynamics on the complete graph with an arbitrarily large population and random networks of finite size. For the fully connected complete graph, the homogeneous mixing condition enables us to use mean-field theory to analyze the opinion evolution of the system.
However, when the number of opinions increases, the number of variables describing the system grows exponentially. 
	To mitigate this, we focus on a special scenario where the largest group of committed agents competes with a motley of committed groups, each of which is smaller than the largest one, while initially, most of uncommitted agents hold one unique opinion. 
This scenario is chosen for its recurrence in diverse societies and its potential for complexity reduction by unifying agents from smaller committed groups into one category.  
Our investigation reveals that when the size of the largest committed group reaches the critical threshold, most of uncommitted agents change their beliefs to this opinion, triggering a phase transition.
	Further, we derive the general formula for the multi-opinion evolution using a recursive approach, enabling investigation into any scenario.
	Finally, we employ agent-based simulations to reveal the opinion evolution and dominance transition in random graphs. 
Our results provide insights into the conditions under which the dominant opinion emerges in a population and the factors that influence these conditions.
\end{abstract}


\IEEEpeerreviewmaketitle

\section{Introduction}
Research on opinion spreading and collective behavior in social systems has spanned over four decades, with significant interest from both mathematical and sociophysics perspectives \cite{degrootReachingConsensus1974, galamApplicationStatisticalPhysics1999, castellanoStatisticalPhysicsSocial2009}. The seminal voter model, introduced by Holley \textit{et al.} \cite{holley_ergodic_1975}, initiated this exploration, wherein actors, holding binary opinions (-1 and +1), adopt the opinion of their randomly chosen neighbors at each step. Subsequent models such as the Sznajd model \cite{sznajd-weron_opinion_2000}, and majority-rule model \cite{galam_minority_2002} have been proposed to investigate binary opinion competition and language evolution. Later, the Naming Game model \cite{baronchelliIndepthAnalysisNaming2008, baronchelliRoleFeedbackBroadcasting2011} has been developed to study language emergence and evolution, allowing for some agents to hold both opinions simultaneously \cite{lipowska_emergence_2022}. Recently, evolutionary game models have emerged to elucidate social influencing from the perspective of cooperative behavior \cite{li_open_2023, feng_evolutionary_2024}.

\begin{comment}
reaction-diffusion model\cite{ vespignaniModellingDynamicalProcesses2012}, network \cite{  anghelCompetitionDrivenNetworkDynamics2004}
\cite{dongSurveyFusionProcess2018, toscaniOpinionModelingSocial2018}
opinion polarization and time to reach \cite{bhatNonuniversalOpinionDynamics2019}
 bistable \cite{baumannLaplacianApproachStubborn2020},  
 minority\cite{yangDynamicalSystemModel2021, iacopiniGroupInteractionsModulate2022} 
Agent-based models and statistical physics provide powerful tools for studying the opinion dynamics and social influence, often modeled by dyadic agent interactions \cite{baronchelliSharpTransitionShared2006, baronchelliIndepthAnalysisNaming2008, liOpinionDynamicsModel2020, yeCollectivePatternsSocial2021}. 
naming game committed agents \cite{, xieEvolutionOpinionsSocial2012, vermaImpactCompetingZealots2014, niuImpactVariableCommitment2017, centolaExperimentalEvidenceTipping2018}
\end{comment}

Here, we employ the Naming Game (NG) model to study the opinion dynamics under various scenarios. Introduced as a linguistic evolution model, the NG was initially used as a model for the formation of a vocabulary from different observations, and it demonstrated how a population of agents can collectively converge to a single unique word for labeling different objects or observations in their environment \cite{baronchelliIndepthAnalysisNaming2008, baronchelliSharpTransitionShared2006}. 
Later, it has been used as a mathematical model for the dynamics of social influence, which describes the evolution of competing opinions through the dyadic interactions between agents. 
Various approaches have been proposed to investigate the evolution and dynamics within the NG model, including mean-field theory \cite{xieSocialConsensusInfluence2011, xieEvolutionOpinionsSocial2012, castelloConsensusOrderingLanguage2009}, agent-based models \cite{ilyinsky_agent-based_2021}, and Bayesian theory \cite{marchettiBayesianApproachNaming2020, marchetti_birds-eye_2020, marchetti_role_2021}.
A number of studies have examined the spread and evolution of opinions on regular lattices \cite{castelloConsensusOrderingLanguage2009}, as well as on diverse complex networks, including random graphs \cite{dallasta_nonequilibrium_2006, luNamingGameSocial2009}, small-world networks \cite{dallasta_agreement_2006, liuNamingGameSmallworld2009}, and scale-free networks \cite{tang_role_2007}.

\begin{comment}
Yet many of them focus on the models with two competing opinions. To gain a general understanding of this model, the scenario with multiple opinions deserves more attention. In such systems, agents can hold a variety of opinions, and the dynamics of opinion evolution can be more complex and diverse than in the two-opinion scenario. Therefore, our study extends the traditional two-opinion NG to the multi-opinion model.
\end{comment}

Furthermore, recent research has also been conducted to understand the NG model in the presence of committed agents \cite{ xieSocialConsensusInfluence2011, xieEvolutionOpinionsSocial2012, thompsonPropensityStickinessNaming2014, doyleSocialConsensusTipping2016}.
When individuals encounter multiple discrete choices or opinions, some may follow the choices of their peers or acquaintances. However, other individuals in the system may advocate a single opinion and refuse to consider any others, to which we refer as committed agents or zealots \cite{mobilia_does_2003, mobiliaRoleZealotryVoter2007}. The presence of zealotry strongly biases the evolution of the opinions towards those held by the committed minorities \cite{galamRoleInflexibleMinorities2007}. Even the presence of one group with committed agents of modest size may convert most uncommitted agents to adopting the opinion of committed agents \cite{cardillo_critical_2020, glassSocialInfluenceCompeting2021}, and such phenomena have been observed in real social systems and experiments \cite{ efferson_promise_2020, andreoni_predicting_2021, yeCollectivePatternsSocial2021}. 

In this study, we focus on the Naming Game with multiple competing opinions and explore how committed members influence opinion evolution. Given the presence of mixed states that involve more than a single opinion, monitoring the state of the system with $m$ distinct single opinions becomes extremely challenging, as there are $2^m-1$ possible combinations of opinions which are proportional to the number of state variables needed to describe the system evolution. Such exponential growth of state variables makes this problem intractable even for the case with the number of opinions, $m$, larger than $10$.

There are a limited number of studies discussing the effects of committed minorities on the evolution of the system and on possible tipping points in multi-opinion Naming Games \cite{waagenEffectZealotryHighdimensional2015, pickeringAnalysisHighdimensionalNaming2016}.
For some special scenarios, one may reduce the system complexity by inspecting symmetry and making appropriate approximations \cite{waagenEffectZealotryHighdimensional2015}. We adapt this approach to investigate the influence of committed agents and phase transition in the quasi-symmetric setup. However, the approximation might fail if no symmetry is preserved. Our strategy is to focus on the key features of the system. Since the system state is determined by the density evolution of each single opinion, it is not necessary to distinguish or record all mixed states. Instead, one only needs to keep track of the density distribution and spreading probability of each single opinion. By anonymizing mixed states, the number of states to be monitored is reduced, making the analysis of the system more manageable. This approach is general and can be applied to a wide range of scenarios. 

The main contributions of this article can be summarized as follows.
\begin{enumerate}
    \item We design the multi-opinion Naming Game model to study the scenario where the group of the largest committed size competes with other smaller committed groups on a complete graph, and we identify the critical transition for the largest group to dominate the system. 
    \item Two special scenarios are constructed to approximate the opinion evolution and identify tipping points for the system under arbitrary configurations, which significantly reduces system complexity and facilitates the analysis of opinion competition and dominance transition.
    \item We observe that the groups of smaller committed sizes can either promote or hinder the opinion of the largest committed size to dominate the system, depending on the number and distribution of committed agents among the small groups.
    \item 
    A recursive approach for the discrete-time NG dynamics on a complete graph is derived, yielding results consistent with the mean-field theory. This method enables a precise description of the system's behavior under any initial configuration.
    \item Agent-based models are employed to simulate the discrete-time dynamics across three types of finite-size networks of real-world characteristics, illustrating the ``divide and conquer" phenomena.
\end{enumerate}

The rest of the paper is organized as follows. Section II provides an overview of the interaction mechanism of the Naming Game and its variants, as well as its dynamical evolution from the perspective of mean-field theory.  
	Section III focuses on the original model on complete graphs and discusses critical transitions for three designed scenarios.
	Section IV presents a recursive approach for the listener-only variant of the Naming Game on complete graphs.
	In Section V, we employ the agent-based model to simulate the original Naming Game model on complex networks. 
 Finally, we summarize our findings and discuss potential avenues for future research.

\section{Model Description and Mean-Field Approximation}
In the Naming Game (NG) model \cite{baronchelliIndepthAnalysisNaming2008,baronchelliSharpTransitionShared2006,castelloConsensusOrderingLanguage2009} with several distinct opinions, each agent holds a subset of opinions that defines its state. This state may change because of this agent's interaction with other agents when it acts as a speaker or listener. 

\begin{figure}[h]
    \centering
    \subfloat[Interaction between uncommitted speaker and uncommitted listener]{
    \includegraphics[width=0.8\linewidth]{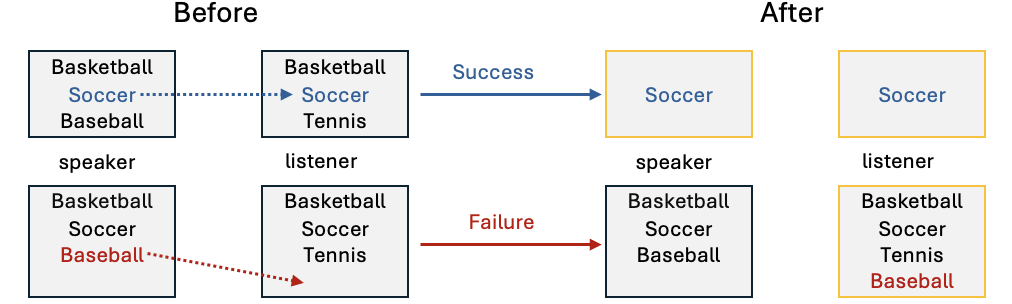}}
    
    \subfloat[Interaction between committed speaker and uncommitted listener]{
    \includegraphics[width=0.80\linewidth]{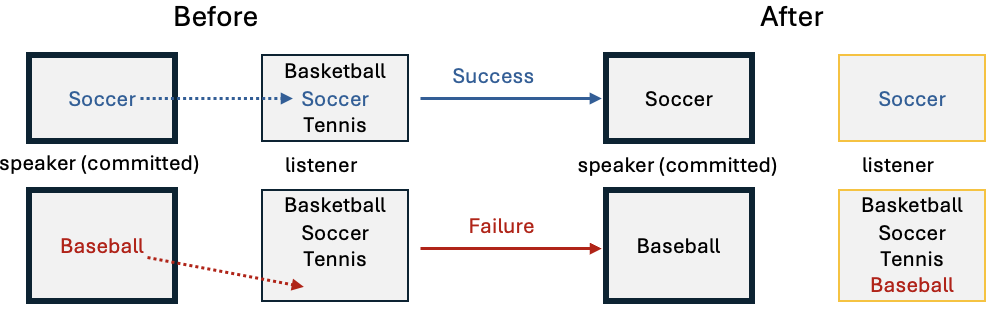}}
    
    \subfloat[Interaction between uncommitted speaker and committed listener]{
    \includegraphics[width=0.80\linewidth]{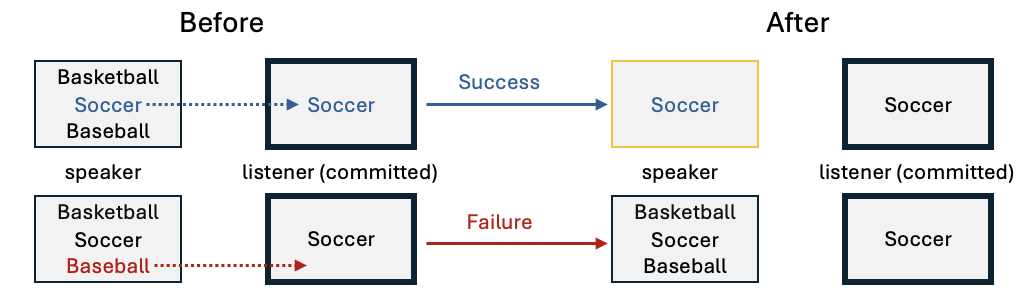}}
    
    \subfloat[Interaction between committed speaker and committed listener]{
    \includegraphics[width=0.80\linewidth]{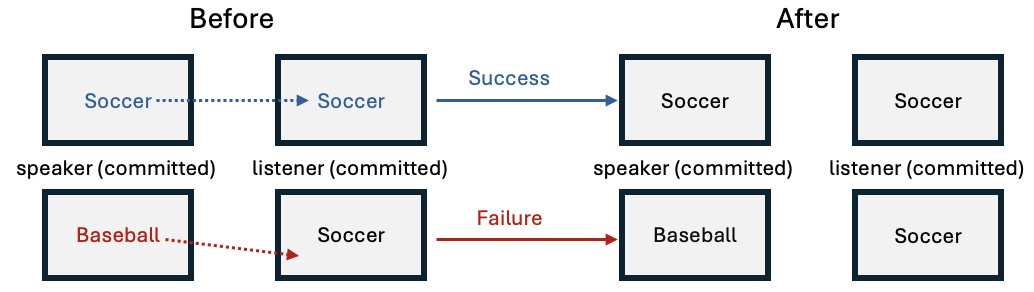}}
    \caption{\textbf{Illustration of model dynamics (Original version). Agents hold one or multiple opinions on the question ``the most popular sport in the world", and may update their opinions after an interaction (indicated by the yellow border of the opinion box).} (a) An uncommitted speaker sends one of its three opinions randomly (``soccer" in the example) to an uncommitted neighbor (listener). If the listener already holds this opinion, both agents retain only this sent opinion (``soccer") as their new state, which is considered a success towards consensus. Otherwise, the listener adds the sent opinion to its state, resulting in a failure. (b) A committed speaker sends the only opinion to an uncommitted listener. Only the listener may change its status depending on whether the consensus is reached. (c) An uncommitted speaker communicates with a committed listener. Similar to (b), only the speaker may change its status. (d) Both speaker and listener are committed to a single opinion. Their statuses are not updated regardless of whether it is a success or failure. }
    \label{fig:diagram}
\end{figure}

	For the original version of NG dynamics, at each NG state, a randomly selected agent acts as a speaker. This speaker randomly chooses an opinion from its opinion state and sends it to a randomly selected neighbor, who then becomes the listener. If the listener already has the sent opinion in its opinion state, both speaker and listener retain only this opinion, otherwise, the listener adds it to its opinion state. 
	There is a special type of agent whose opinion state contains only one opinion, and it holds its opinion unchanged during the entire dynamics. Such agents are immune to any influence but can spread their opinions to their neighbors when acting as speakers. We refer to them as committed agents or zealots. The model mechanism is summarized in Fig.~\ref{fig:diagram}.
	In addition to this original model, there are two variants, which limit changes to only one of the two interacting roles, named the ``listener-only" and ``speaker-only" versions.
	For the ``listener-only" type, only the opinion state of listeners can be modified. 
	In this paper, we focus on the original NG model and its ``listener-only'' variant. 

	First, we investigate the opinion dynamics on the complete graph, where mean-field theory can be applied to systematically study the evolution of opinion states. 
	For the general scenario with $m$ unique single opinions,  an uncommitted agent can hold one of $M=2^m-1$ opinion states at each stage. For instance, when $m=3$, the possible opinion states are $A$, $B$, $C$, $AB$, $AC$, $BC$, and $ABC$.
 Under the condition of homogeneous mixing, the mean-field differential equations are written as
	\begin{equation}
 \begin{split}
		\dv[]{x_k}{t} =& \sum _{i=1}^{M} \sum _{j=1}^{M} U_{ij} ^{(k)} x_i x_j + \sum _{i=1}^{M} \sum _{j=1}^{m} V_{ij} ^{(k)} x_i P_j   \\
  &+ \sum _{i=1}^{m} \sum _{j=1}^{M} W_{ij} ^{(k)} P_i x_j .\label{eq:general_mft}
 \end{split}
	\end{equation} 
 
This equation describes the changes in the density of uncommitted agents holding different opinion states as well as the interactions between the uncommitted agents and committed agents.
The density $x_i (i=1, 2,..., m)$ represents the fraction of uncommitted agents holding the single opinion state $i$, and the density $x_i$ ($i=m+1, m+2, ..., M$) represents the fraction of agents holding the mixed opinion state $i$. $P_i (i =1, 2, ..., m)$ is the density of zealots committed to the single opinion $i$, which remains constant over time. The matrices $U$, $V$, and $W$ contain the coefficients determined by the interaction mechanism, and they differ for the three versions of the interaction rules. Specifically, $U_{ij}^{(k)}$ is the probability that the interaction between the uncommitted speaker with the opinion state $i$ and the uncommitted listener with $j$ gives rise to the opinion state $k$. $V_{ij}^{(k)}$ is the probability that results in the speaker adopting the opinion state $k$ for the interaction between the uncommitted speaker holding the opinion state $i$ and the committed listener with $j$. Similarly, $W_{ij}^{(k)}$ is the probability that results in the listener adopting the opinion state $k$ for the interaction between the committed speaker holding the opinion state $i$ and the uncommitted listener with $j$. 
The densities $x_i$ and $P_i$ must sum up to 1, so we have $\sum _{i=1}^{M} x_i + \sum _{i=1}^{m} P_i = 1$.

\begin{figure*}[h]
    \centering
    \subfloat[dominance transition of $A$]{
    \includegraphics[width=0.45\linewidth]{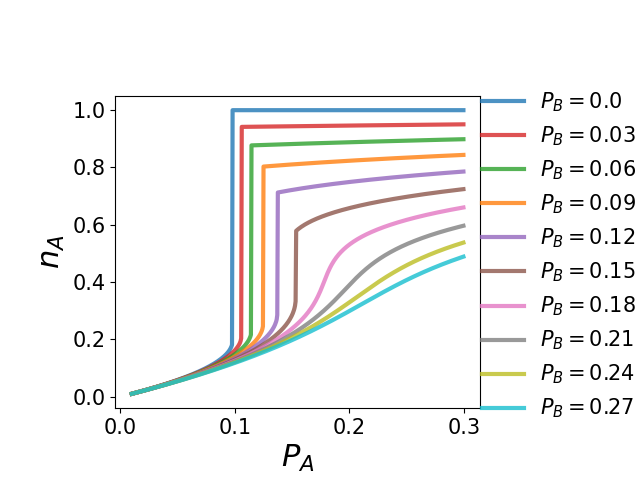}  }
    \subfloat[tipping point $P_A^{(c)}$]{
    \includegraphics[width=0.45\linewidth]{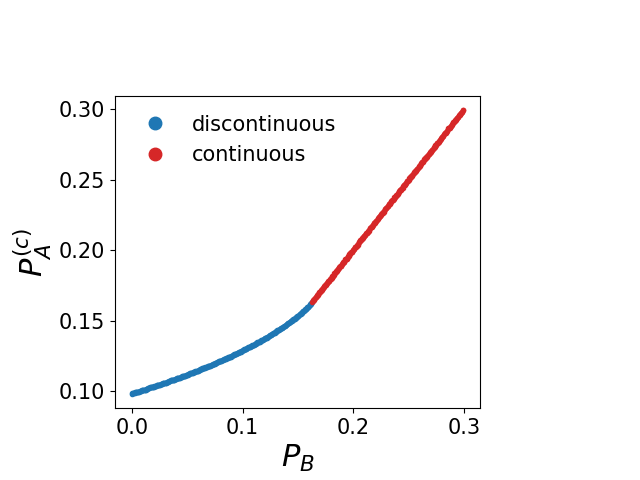}}
    \caption{\textbf{Phase transition from $B$ dominance to $A$ dominance and the corresponding tipping points for two-opinion scenario ($m=2$).} All uncommitted agents support $B$ initially. (a) The stable density of agents with opinion $A$, $n_A$, changes as a function of the committed fraction $P_A$ for different values of $P_B$. As $P_A$ increases, the system is dominated by $A$. (b) The critical point $P_A^{(c)}$ changes with $P_B$. The increase in $P_B$ raises the value of the critical points for $A$ to dominate. The blue dots represent the discontinuous transition of $n_A$ versus  $P_A$, while the red ones represent the continuous change. }
    \label{fig:F1}
\end{figure*}

For the system with a small number of single opinions, $m$, the numerical integration of the mean-field differential equation, Eq.~\myref{eq:general_mft}, can be performed to obtain the density evolution of each opinion state in the NG model. However, as the number of all opinion states, $M$, which includes both single and mixed opinions, increases exponentially with $m$, performing direct numerical simulations becomes computationally infeasible and impractical for large values of $m$. 

\section{Original version}

First, the original version of NG dynamics is analyzed using mean-field differential equations, with a focus on the density evolution of each opinion state in the presence of committed minorities. This section includes the study of three scenarios varying in complexity, the first with two single opinions, the second with three single opinions, and the third with $m$ single opinions in general.

	\subsection{The two-opinion scenario}
 
\begin{figure*}[h]
    \centering
    \subfloat[the fraction of agents supporting $A$]{
    \includegraphics[width=0.45\linewidth]{nA_PA_N=3_heatmap}}
    \subfloat[the fraction of agents supporting $B$]{
    \includegraphics[width=0.45\linewidth]{nB_PA_N=3_heatmap}}
    \vspace{-3mm}
    \newline
    \subfloat[the fraction of agents supporting $C$]{
    \includegraphics[width=0.45\linewidth]{nC_PA_N=3_heatmap}}
    \subfloat[tipping point $P_A^{(c)}$]{
    \includegraphics[width=0.45\linewidth]{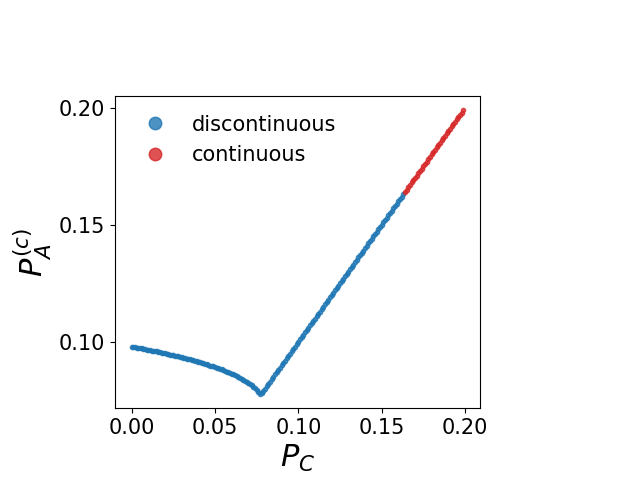}}
    \caption{\textbf{Phase transition from $B$ or $C$ dominance to $A$ dominance and the corresponding tipping points for three-opinion scenario ($m=3$).} All uncommitted agents support $B$ initially. (a) The stable fraction of agents with opinion $A$ changes as a function of the committed fractions $P_A$ and $P_C$. Similarly, (b) and (c) show the change of $n_B$ and $n_c$, respectively. As $P_A$ increases, the system is dominated by $A$, including continuous and discontinuous transitions. When $P_A$ is below the tipping point $P_A^{(c)}$, the system is dominated by $B$ for small $P_{C}$ ($<0.1$) and dominated by $C$ for large $P_{C}$. (d) The critical point $P_A^{(c)}$ changes with $P_C$. As $P_C$ increases, the transition $n_A$ versus  $P_A$ changes from the discontinuous transition (blue dots) to the continuous transition (red dots).}
    \label{fig:F2}
\end{figure*}
	In the scenario of $m=2$, there are two opinions, $A$ and $B$, in the system competing against each other. Eq.~\myref{eq:general_mft} reduces to two mean-field equations, 
\begin{equation}
\begin{split}
	\dv[]{x_A}{t} &= -x_A x_B + x_{AB}^2 + x_{AB} x_A + \frac{3}{2}P_A x_{AB} - P_B x_A\\
	\dv[]{x_B}{t} &= -x_A x_B + x_{AB}^2 + x_{AB} x_B + \frac{3}{2}P_B x_{AB} - P_A x_B 
\end{split}. \label{eq:two_opinion}
\end{equation}
By definition, $x_A + x_B + x_{AB} + P_A + P_B =1$.
Together with Eq.~\myref{eq:two_opinion}, the two-opinion model can be analytically and numerically solved.
Such a system can exhibit rich dynamics, including saddle-node bifurcation, indicating that it may have multiple stable equilibria \cite{xieEvolutionOpinionsSocial2012, verma_impact_2014}. Additionally, the dominance of the system is primarily determined by the committed sizes of two competing opinions \cite{xieEvolutionOpinionsSocial2012}.

Here, we are interested in the scenario in which one opinion (let us say $A$) has a higher fraction of committed agents than the other opinion, $B$, but the latter is initially supported by all uncommitted agents, making it the majority opinion. However, committed agents of opinion $A$ can assimilate uncommitted agents, thus causing opinion $A$ to eventually become the majority opinion. Previous studies \cite{xieSocialConsensusInfluence2011, xieEvolutionOpinionsSocial2012} have shown that there exists a minimal fraction of committed agents, denoted by $P_A^{(c)}$, which is required for a fast phase transition of the dominant opinion from $B$ to $A$. Below this threshold, the waiting time for such a transition grows exponentially with the number of agents, making it infeasible to observe in practical cases.

To understand the final dominant state of the system, a new variable, $n_i$, is introduced, which represents the total fraction of agents holding opinion $i$ in equilibrium. This fraction includes both the committed and uncommitted agents that support opinion $i$, $n_i = x_i^{(s)} + P_i$, whereas for mixed opinion states, $n_i$ only accounts for the uncommitted agents, $n_i = x_i^{(s)}$, because committed agents only advocate their single opinions. 
Previous studies \cite{xieSocialConsensusInfluence2011} have shown that in the absence of committed agents advocating opinion $B$ ($P_B = 0$, $P_A>0$), a minimal fraction of committed agents advocating opinion $A$ ($P_A^{(c)}$) of approximately $0.098$ is required to trigger a fast transition from the majority opinion $B$ to $A$.
As Fig.~\ref{fig:F1} shows, when both committed groups, $A$ and $B$, are present, there are two types of transitions, the discontinuous transition and the continuous one, which may occur depending on their committed fractions. They are separated by the point $(P^{(c)}, P^{(c)}) \approx (0.162, 0.162)$ \cite{xieEvolutionOpinionsSocial2012}. For $P_B>P^{(c)}$, the fraction of agents holding opinion $A$ increases continuously with $P_A$, and the critical points lie on the line $P_A^{(c)} =P_B$.

\subsection{Three-opinion scenario}

A slightly more complex system arises with three opinions: A, B, and C, with two opinions A and C committed by two minor fractions of committed agents and, initially, the majority of agents are uncommitted and they all hold opinion B. 
We ask a similar question as in the previous example. For the scenario of $P_A > P_C$, to enable opinion $A$ to dominate the system, what is the minimal fraction of committed agents,  $P_A^{(c)}$, and how does this threshold depend on the committed fraction of the opinion $C$? 
According to Eq.~\myref{eq:general_mft}, the evolution of each state variable can be numerically integrated. 

Depending on the fractions of agents committed to opinions $A$ and $C$, the system can be dominated by any of three opinions. Observed from Fig.~\ref{fig:F2}, for small values of $P_C$ ($<0.06$), the system exhibits a discontinuous transition from being dominated by $B$ to $A$ dominance when the committed fraction $P_A$ is above the critical point. In contrast, for large values of $P_C$, the system undergoes a continuous transition, where opinion $A$ wins the competition against $C$ by increasing $P_A$ to the critical point. Additionally, the relationship between the critical point $P_A^{(c)}$ and $P_C$ is non-monotonic, as shown in Fig.~\ref{fig:F2}d. As $P_C$ increases, $P_A^{(c)}$ decreases first with the transition being discontinuous. These observations indicate that increasing the population committed to $C$ speeds up the spread of opinion $A$ to the majority of uncommitted agents, as long as $P_C$ is smaller than a certain value ($P_C \approx 0.077$ at the lowest point in Fig.~\ref{fig:F2}b). Otherwise, $P_A^{(c)}$ increases linearly with $P_C$, signaling a change of relationship between opinions $A$ and $C$ from collaboration to competition. Unlike the previous two-opinion scenario, this one includes both the discontinuous transition and the continuous one. It is noteworthy that the critical point separating the two types of transitions remains the same as in the two-opinion scenario.

\subsection{The general scenario -- multi-opinion model}

\begin{figure*}[h]
    \centering
    \subfloat[the fraction of supporting $A$]{
    \includegraphics[width=0.30\linewidth]{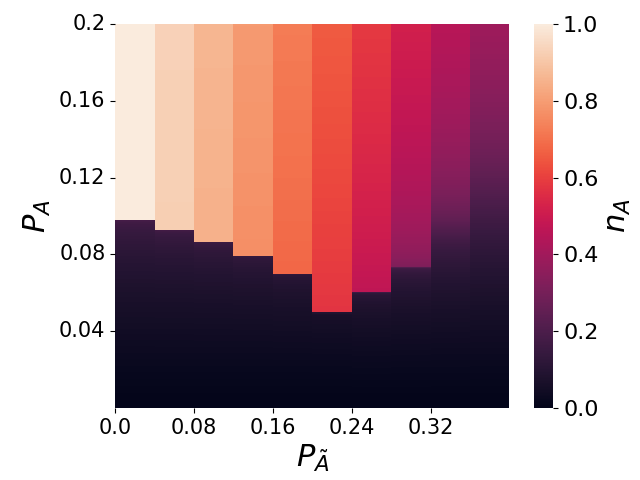}}
    \subfloat[the fraction of supporting $B$]{
    \includegraphics[width=0.30\linewidth]{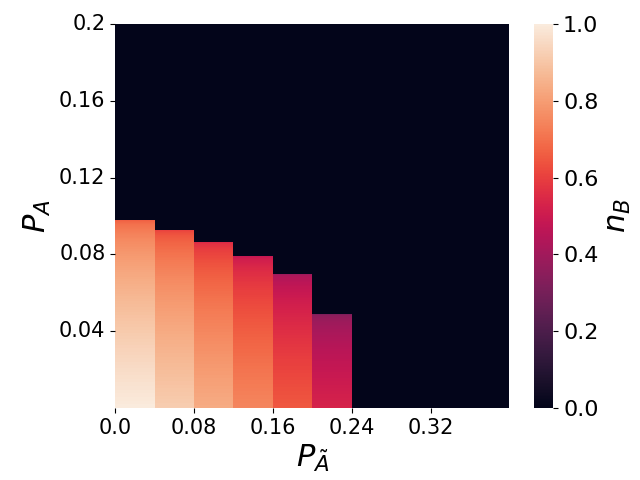}}
    \subfloat[the fraction of supporting $C$]{
    \includegraphics[width=0.30\linewidth]{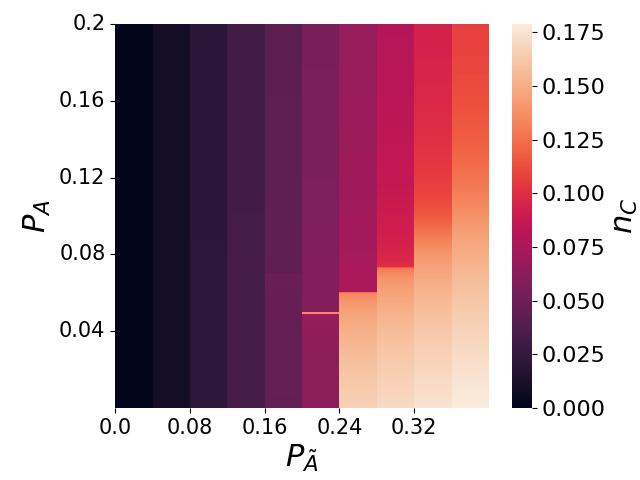}}
    \caption{\textbf{The density of agents supporting single opinions at the steady state in scenario $S_1$.} The number of opinions is set as $m=6$. The fractions of agents holding opinions $A$, $B$, and $C$ (representing any single opinion $C_1$, $C_2$, $C_3$, or $C_4$ in group $\tilde{A}$) are shown in subfigures (a) -- (c), respectively. As $P_A$ increases, the system is dominated by $A$, including continuous and discontinuous transitions. When $P_A$ is below the tipping point $P_A^{(c)}$, the system is dominated by $B$ for small $P_{\tilde{A}}$ ($<0.2$) and dominated by the unified group $\tilde{A}$ for large $P_{\tilde{A}}$. }
    \label{fig:F3}
\end{figure*}

\begin{figure*}[h]
    \centering
    \subfloat[critical point changes with $p_0$]{
    \includegraphics[width=0.45\linewidth]{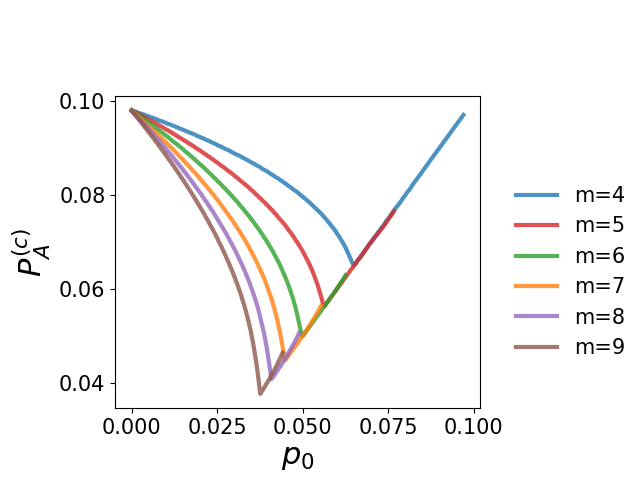}}
    \subfloat[critical point changes with $P_{\tilde{A}}$]{
    \includegraphics[width=0.45\linewidth]{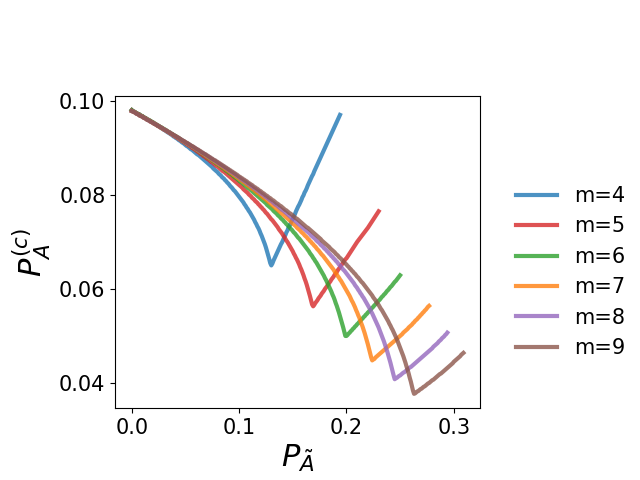}}
    \caption{\textbf{The phase transition and the critical points in scenario $S_1$.} For different values of $m$ ($m=4, 5, 6, 7, 8, 9$), the critical point $p_A^{(c)}$ changes with (a) $p_0$ and (b) $P_{\tilde{A}}$ ($P_{\tilde{A}} = (m-2) p_0$). The initial decrease of $P_A^{(c)}$ with $p_0$ (or $P_{\tilde{A}}$) indicates that the minority group facilitates the dominance of $A$, corresponding to the discontinuous transition. The linear increasing regime suggests the competition between $A$ and the unified group $\tilde{A}$, corresponding to the continuous transition.}
    \label{fig:F4}
\end{figure*}

\begin{figure*}[h!]
    \centering
    \subfloat[critical point changes with $max(P_i)$]{
    \includegraphics[width=0.45\linewidth]{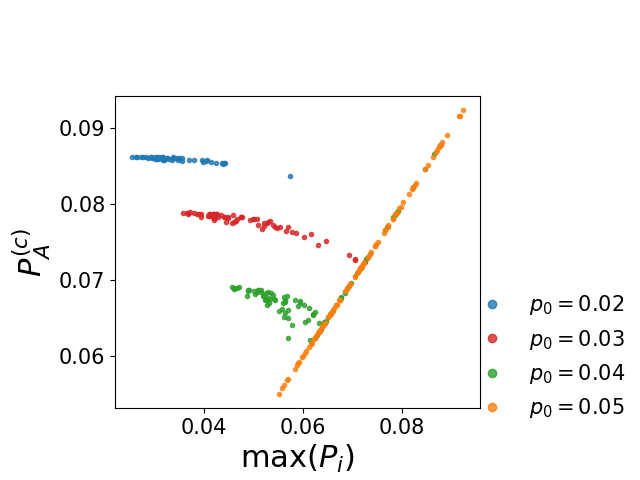}}
    \subfloat[critical point changes with $SD(P_i)$]{
    \includegraphics[width=0.45\linewidth]{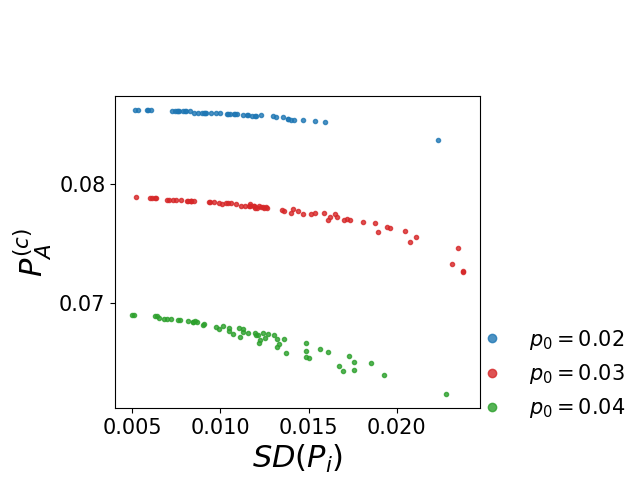}}
    \caption{\textbf{The critical point for $A$ to dominate in scenario $S_0$.} The number of opinions is set as $m=6$. (a) For different values of the minority committed fraction, $p_0=0.03, 0.04, 0.05, 0.06$, $P_A^{(c)}$ changes with the maximum of $P_i$ in group $\tilde{A}$ with an initial decrease followed by a linear increase.  (b) $P_A^{(c)}$ changes with the standard deviation (SD) of $P_i$, which only includes the data of the decreasing regime in (a). This indicates that stronger opponents within the unified group $\tilde{A}$ may lower the critical values for $A$ to dominate, provided that this unified group is not the largest opponent.}
    \label{fig:F5}
\end{figure*}

For the general scenario with $m$ single opinions ($A$, $B$, $C_1$, $C_2$, $C_3$, ..., $C_{m-2}$), it is of interest to understand the impact of committed agents on the majority of uncommitted agents and potential for one single opinion to dominate over other competitors. Consider a scenario where most of uncommitted agents support a single opinion, denoted as $B$, while the remaining agents are committed to $m-1$ single opinions. Among these $m-1$ opinions, the one with the largest committed fraction, denoted as $A$, has the ability to reverse the majority of uncommitted agents from supporting $B$ to supporting $A$. The question then arises as to the minimum fraction of committed agents, $P_A^{(c)}$, required for such a transition to occur. To streamline the analysis, the committed agents supporting opinions other than $A$ are grouped into a single category, referred to as $\tilde{A}$, with a combined committed fraction of $P_{\tilde{A}}$. This simplification is justified as none of the single opinions in the group $\tilde{A}$ can prevail in the competition with a larger committed group $A$. However, the number of competing opinions in the group $\tilde{A}$, $m-2$, their total committed fraction, $P_{\tilde{A}}$, and the allocation of these committed agents, $P_i$, may all potentially affect the critical point, $P_A^{(c)}$.

Hence, we investigate the impact of such factors on the dominance transition of opinion dynamics by constructing three different scenarios for allocating committed agents within the group $\tilde{A}$.

\begin{enumerate}
	\item \textbf{Scenario $S_0$: randomly distributed.} The committed fraction, $P_i$, of each single opinion in group $\tilde{A}$ is generated by a truncated Gaussian distribution with a mean of $p_0 = P_{\tilde{A}}/(m-2)$, a predefined standard deviation $\sigma =0.02$ and a restricted interval $[0, P_{\tilde{A}}]$. One should note that the actual standard deviation can differ from the predefined value as shown in Fig.~\ref{fig:F5}. This distribution allows for any value between $0$ and $P_{\tilde{A}}$, though subject to a constraint that their sum totals $P_{\tilde{A}}$.

	\item \textbf{Scenario $S_1$: perfectly symmetric.} $m-2$ opinions in the group $\tilde{A}$ share the equal fraction of committed agents, $P_i = p_0$. The quantity, $p_0$, in the later context also refers to the average committed fractions of agents advocating any single opinion in the group $\tilde{A}$.

	\item \textbf{Scenario $S_2$: extremely polarized.} In contrast to scenario $S_1$, we maximize the deviation of $P_i$ in group $\tilde{A}$ to establish a highly uneven distribution of committed fractions.
		Provided that the single opinion $A$ has the largest committed fraction in the system, the largest committed fraction in the group $\tilde{A}$ should be smaller than $P_A$. To set up the numerical simulation, we choose $\max\{P_i\}=p_1 = P_A - 10^{-3}$ and maximize the number of opinions with the committed fraction $p_1$, which is $n_1=\left \lfloor{P_{\tilde{A}}}/p_1\right \rfloor$. The remaining committed agents, $p_2 = P_{\tilde{A}} - n_1 p_1 (< p_1)$, are assigned to another single opinion. In this scenario, there are $m-n_1-3$ ($\ge 0$) single opinions in group $\tilde{A}$ without any committed followers. Within group $\tilde{A}$, $P_{i}$ can take three values, $p_1$, $p_2$, and $0$. As there are no uncommitted agents assigned to group $\tilde{A}$, some single opinions may end up with no supporters. To compare with scenarios $S_0$ and $S_1$, the number of single opinions is still considered as $m$. 
\end{enumerate}

\begin{figure*}[h]
    \centering
    \subfloat[$m=4$]{
    \includegraphics[width=0.30\linewidth]{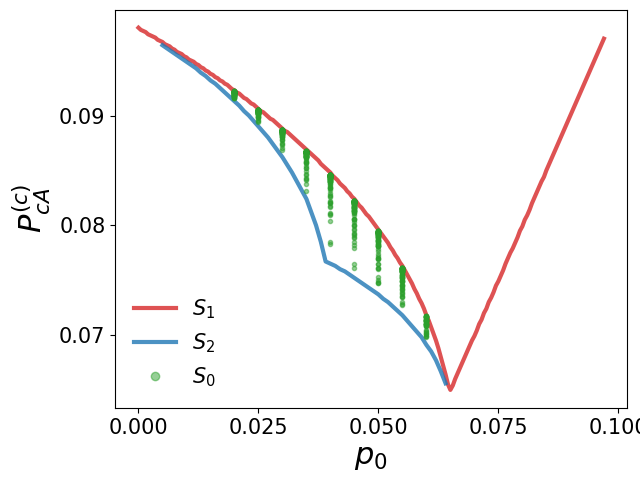}}
    \subfloat[$m=5$]{
    \includegraphics[width=0.30\linewidth]{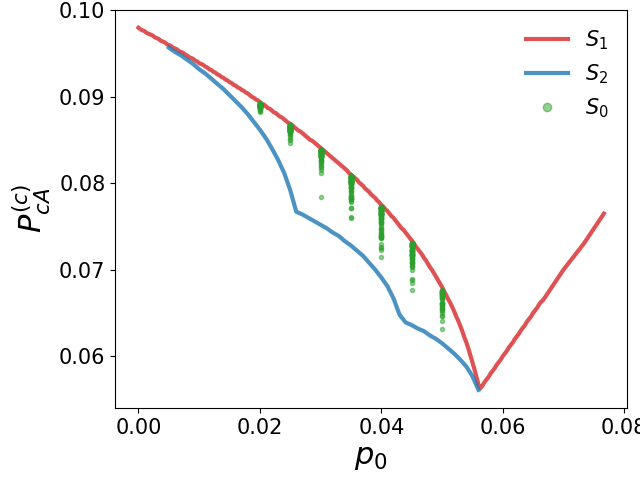}}
    \subfloat[$m=6$]{
    \includegraphics[width=0.30\linewidth]{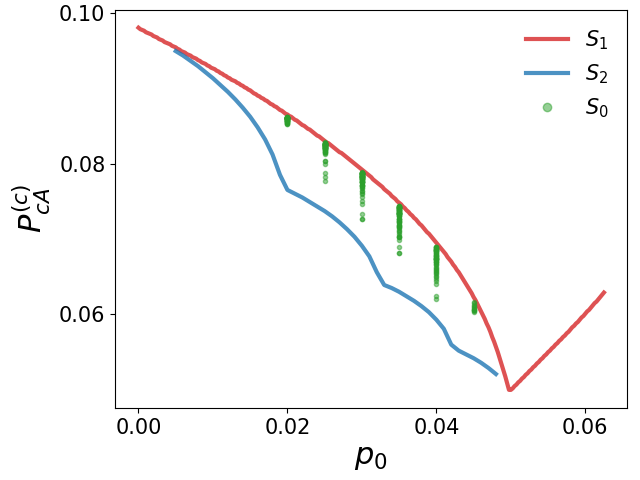}}
    \caption{\textbf{The critical point $P_A^{(c)}$ changes with $p_0$ in three scenarios $S_0$, $S_1$, and $S_2$.} The number of opinions is set as (a) $m=4$, (b) $m=5$, (c) $m=6$. For scenario $S_0$, only the data where $P_A^{(c)}$ is along the decreasing branch with $\max(P_i)$ in Fig.~\ref{fig:F5} is included. The tipping points in scenarios $S_1$ and $S_2$ provide the upper and lower bound for $S_0$. }
    \label{fig:F6}
\end{figure*}

\begin{figure*}[h]
    \centering
    \subfloat[$p_0=0.02$]{
    \includegraphics[width=0.30\linewidth]{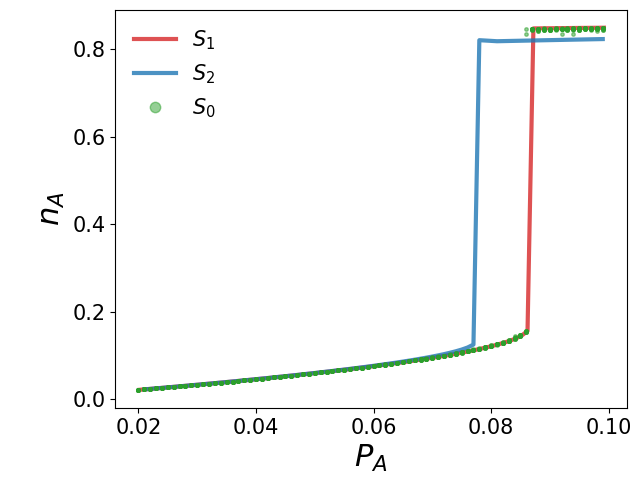}}
    \subfloat[$p_0=0.04$]{
    \includegraphics[width=0.30\linewidth]{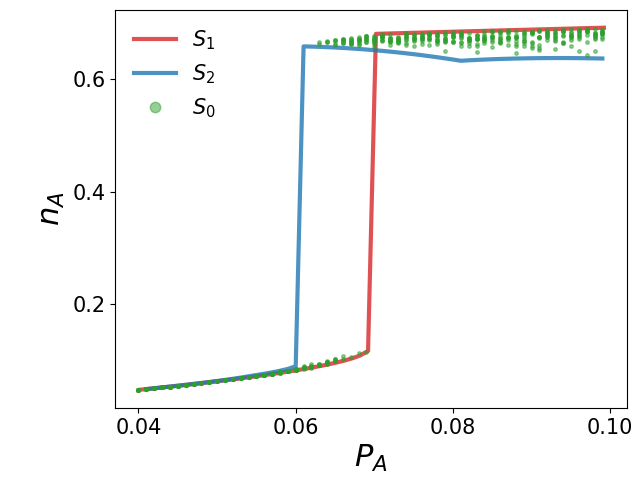}}
    \subfloat[$p_0=0.06$]{
    \includegraphics[width=0.30\linewidth]{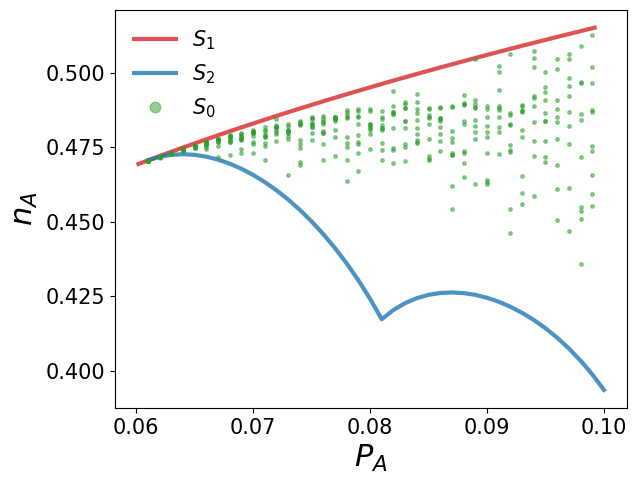}}
    \caption{\textbf{The steady state $n_A$ changes with $P_A$ in three scenarios $S_0$, $S_1$, and $S_2$ with $m=6$ opinions.} The average committed fraction in group $\tilde{A}$ is set as (a) $p_0=0.02$; (b) $p_0=0.04$; (c) $p_0=0.06$. The stable fraction in scenario $S_0$ can be approximated by $S_1$ and $S_2$. }
    \label{fig:F7}
\end{figure*}

The mean-field equations \myref{eq:general_mft} can be directly integrated to analyze the opinion dynamics for a system with a limited number of single opinions. However, for a system with many opinions $m$, this method becomes computationally infeasible because the number of variables, $M$, increases exponentially with $m$. To overcome this challenge, simpler scenarios with symmetry are considered, as described in scenarios $S_1$ and $S_2$.  The simplified structures of scenarios $S_1$ and $S_2$ allow for a more efficient and manageable study of the critical transition in comparison to direct numerical integration for scenario $S_0$ with random initial configurations.
In scenario $S_1$, a collection of single opinions (denoted as the group $\tilde{A}$) is designed to have an equal fraction of committed agents with no uncommitted supporters. Under the assumption of homogeneous mixing in a complete graph, the fraction of supporters for these opinions is expected to evolve in the same fashion. Consequently, the number of state variables to be monitored is reduced from $2^m -1$ to $4m-5$. For example, when $m=5$ where single opinions are $A$, $B$, $C_1$, $C_2$, $C_3$. Opinions $C_1$, $C_2$, and $C_3$ are assigned the same fraction of committed agents, so the fraction of uncommitted agents they can assimilate to themselves is expected to be the same by symmetry. Further, some mixed opinion states, such as $C_1C_2$, $C_1C_3$, and $C_2C_3$, or $AC_1$, $AC_2$, and $AC_3$ also have the same uncommitted supporters as time progresses. This results in a reduction in the number of state variables that need to be monitored. A similar argument also applies to scenario $S_2$ as some of the opinions in group $\tilde{A}$ have the same fraction of committed agents.

Next, we focus on the evolution of the fraction of agents supporting opinion $A$, which is assigned the largest committed fraction, and explore the critical transition where this opinion assimilates most of the uncommitted individuals across the three scenarios. 
In scenario $S_1$, for small values of $P_{\tilde{A}}$, the system undergoes a discontinuous transition from $B$ dominance to $A$ dominance as $P_A$ increases (Fig.~\ref{fig:F3}).  Also, as seen Fig.~\ref{fig:F4}, the critical point $P_A^{(c)}$ shows a non-monotonic behavior as $P_{\tilde{A}}$ or $p_0$ increases. The presence of a small committed group plays a key role in the formation of a dominant opinion. Initially, the critical value $P_A^{(c)}$ decreases as the committed fraction $p_0$ of the small groups increases, indicating that as the number of committed individuals in these groups grows, they become more effective in promoting the dominance of opinion $A$. This implies a catalyzing role of small groups for disseminating opinion $A$ to uncommitted agents. The initial decrease in $P_A^{(c)}$ can be attributed to the increased chance for interactions and conversions between the committed individuals in the smaller groups and the uncommitted individuals in the system.
Moreover, the non-monotonic behavior of $P_A^{(c)}$ with increasing $P_{\tilde{A}}$ or $p_0$ also indicates the presence of a threshold effect. Beyond a certain value of $P_{\tilde{A}}$ or $p_0$, the critical value $P_A^{(c)}$ begins to increase, indicating that the positive influence of the smaller committed groups on the dominant opinion's growth reverses. The linear relationship instead shows the competition between opinion $A$ and other opinions with a smaller committed fraction, which can also be confirmed by comparing Fig.~\ref{fig:F3}a and c.

To explore how the value of the tipping point $P_A^{(c)}$ depends on the allocation of committed agents to group $\tilde{A}$, we manipulate the committed fraction $P_i$ while preserving $P_{\tilde{A}}$ in scenario $S_0$. 
Results displayed in Fig.~\ref{fig:F5}a show a non-monotonic behavior of the critical point $P_A^{(c)}$ as a function of the maximum value of $P_i$ in group $\tilde{A}$. The initial decrease of $P_A^{(c)}$ indicates that the presence of a large fraction of committed agents within group $\tilde{A}$ is beneficial for opinion $A$ to be adopted by most of the uncommitted agents compared to the case when the committed agents are equally distributed among the $m-2$ single opinions. 

This conclusion can also be confirmed by observing how $P_A^{(c)}$ changes with the standard deviation of $P_i$ in Fig.~\ref{fig:F5}b. However, it is worth noting that a higher $P_i$ does not always result in a favorable outcome in terms of the dominance of opinion $A$. For opinion $A$ to become dominant, its committed fraction $P_A$ must be greater than any other committed fraction in the group $\tilde{A}$, which explains the linear increase of $P_A^{(c)}$ observed in the results of $p_0=0.05$.
The non-monotonic behavior of the critical value of $P_A^{(c)}$ highlights the importance of considering the effects of different distributions of committed fractions on the overall dynamics of the system, especially the dominance transition.

\begin{figure*}[h]
	\centering
	\subfloat[recursive approach]{
	\includegraphics[width=0.46\textwidth]{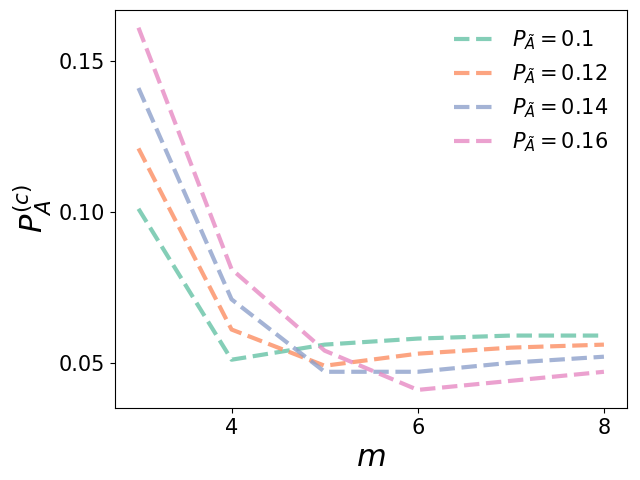}}
	\subfloat[integration of ODEs]{
	\includegraphics[width=0.46\textwidth]{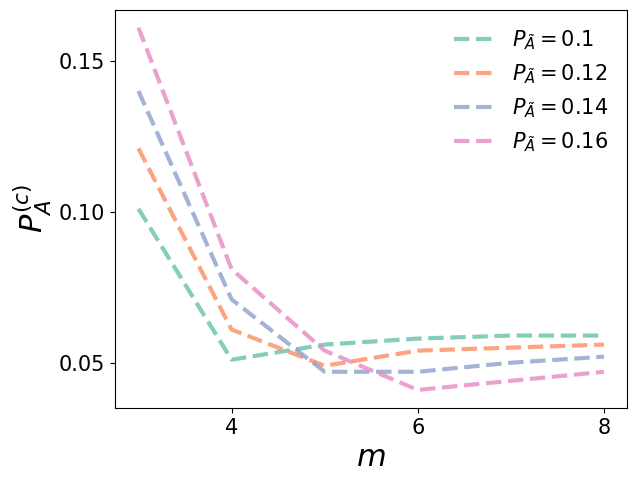}}
	\caption{\textbf{Divide and conquer.} The number of total committed agents in group $\tilde{A}$ is set as $P_{\tilde{A} }= 0.1, 0.12, 0.14, 0.16$. The critical point $P_A^{(c)}$ in scenario $S_1$ is determined by using the recursive approach in (a) and the integration of the differential equations in (b). The critical point, $P_A^{(c)}$, exhibits a non-monotonic relationship with the number of single opinions, $m$. This implies that dividing the committed agents into a moderate number of competing minorities can facilitate the dominance of opinion $A$ among the uncommitted agents in the system. } \label{fig:pAc_N}

\end{figure*}

From the observation in Fig.~\ref{fig:F5}, one may expect that scenario $S_2$ has a smaller critical point $P_A^{(c)}$ than $S_1$ as the standard deviation of committed sizes in group $\tilde{A}$ is maximized. This expectation is confirmed by Fig.~\ref{fig:F6}.
The critical points obtained from two scenarios, $S_1$ and $S_2$, provide the upper and lower bounds for scenario $S_0$, respectively.
Additionally, one can compare the steady states of the three scenarios in Fig.~\ref{fig:F7}. Scenarios $S_1$ and $S_2$ also provide a good approximation for the steady state $n_A$ in scenario $S_0$. It is observed that the critical point $P_A^{(c_1)}$ in scenarios $S_1$ is always greater than $P_A^{(c_2)}$ in scenario $S_2$, and the two critical points $P_A^{(c_1)}$ and $P_A^{(c_2)}$ divide the parameter space into three parts. For values of $P_A$ less than $P_A^{(c_2)}$, scenario $S_1$ yields the lower bound of $n_A$ while $S_2$ provides the upper bound. For $P_A^{(c_2)} < P_A < P_A^{(c_1)}$, both scenarios establish the lower bound. For $P_A > P_A^{(c_1)}$ or the case when there are no critical points, scenario $S_1$ corresponds to the upper limit of $n_A$ while $S_2$ corresponds to the lower limit. 
By investigating scenarios $S_1$ and $S_2$ of symmetric setup, the critical points and the steady states of opinion $A$ with the largest committed fraction in scenario $S_0$ are well estimated.

We now analyze the opinion competition from another perspective. 
The key question is to determine the dynamics of opinion $A$ as it competes against opinions $B$ and $\tilde{A}$. As shown in Fig.~\ref{fig:pAc_N}a, the critical point, $P_A^{(c)}$, in scenario $S_1$ has a non-monotonic relationship with the number of single opinions, $m$. Given a fixed committed fraction, $P_{\tilde{A}}$, as $m$ increases, the individual committed fraction, $p_0$ ($=P_{\tilde{A}} / (m-2)$), in group $\tilde{A}$ decreases, weakening the opposition from this group. 
The initial decrease of $P_A^{(c)}$ reveals the validity of the divide-and-conquer policy, whereby the more opinions split among themselves the committed agents of group $\tilde{A}$, the easier it is for opinion $A$ to dominate uncommitted agents in the system. Reversing this rule reveals that the major obstacle to the opinion $A$ dominance is the small number of opinions in the group $\tilde{A}$. However, if $m$ continues to increase, the critical point $P_A^{(c)}$ also increases, suggesting that opinion $B$ becomes the major threat. In this scenario, a strong opponent, $\tilde{A}$, (large $p_0$) can be helpful for opinion $A$ to dominate the system, thus making group $\tilde{A}$ a friend of opinion $A$, in line with the Heider balance theory rule \cite{heiderAttitudesCognitiveOrganization1946} that states ``The enemy of my enemy is my friend".
This ``divide and conquer" phenomenon has been observed in other systems, such as pathogen infection dynamics, where increasing diversity of host species may either amplify or buffer the disease outbreaks depending on the transmission types \cite{dobson_population_2004}. Additionally, a similar phenomenon has also been reported in the multi-species system, where outside invasions are more likely to succeed as the number of species increases \cite{gjini_towards_2023}.

\begin{comment}
The critical point $P_A^{(c)}$ in scenario $S_0$ can differ depending on the distribution of the committed agents. However, scenarios $S_1$ and $S_2$ serve as an approximation by providing the upper and lower bounds, respectively, for this critical value. Additionally, the symmetry in scenarios $S_1$ and $S_2$ results in identical evolution for opinion states with the same committed fraction in group $\tilde{A}$ under the homogeneous mixing condition. This reduction in complexity allows for a more efficient analysis of the system dynamics, as a satisfactory approximation can be obtained by considering scenarios $S_1$ and $S_2$.  
\end{comment}

	\section{Simplification  by recursive relationship}
 \begin{figure*}[h]
	\centering
	\subfloat[the fraction of supporting $A$ ($B$)]{
	\includegraphics[width=0.45\textwidth]{{xAB_p}}}
	\subfloat[the fraction of supporting $C$]{
	\includegraphics[width=0.45\textwidth]{{xC_p}}}
	\caption{\textbf{Comparison between the recursive approach and the differential equations.} The evolution of the uncommitted fraction for opinions $A$, $B$ (shown in (a)), and $C_1$ (same as $C_2$, $C_3$, $C_4$, thus denoted as $C$) (shown in (b)) are obtained by two methods. The number of opinions is set as $m = 6$, the fraction of agents committed to $A$ is $P_A = 0.1$, and the fraction of agents committed to each minority opinion is $P_C = P_{C_1} = P_{C_2} = P_{C_3} = P_{C_4} =  0.025$. Initially, all the uncommitted agents support opinion $B$, $x_B (t=0) = 0.8$. } \label{fig:compare_recursive_ode}
\end{figure*}

 In the previous section, we explored methods for establishing symmetrical distributions of committed agents and utilizing mean-field frameworks to simplify the continuous-time dynamics, thereby approximating opinion dynamics for scenarios with arbitrary distributions of committed agents.
 In this section, we shift our focus to its discrete-time version and introduce a more general method for reducing system complexity using the recursive approach. This approach enables a more focused examination of the evolution of supporters for single opinions by anonymizing mixed states, facilitating the determination of the dominant opinion in a more efficient manner.

\subsection{Establish recursive relationship}
Since the committed agents define the dominant state in NG dynamics at equilibrium, it is sufficient to focus on only the density evolution of supporters for single opinions. 
We introduce a quantity $Q_i^{(t)}$, which represents the probability of a single opinion $i$ being communicated at step $t$ from the population \cite{waagenEffectZealotryHighdimensional2015}, and we establish an iteration function for the opinion density at step $t$ based on the state at step $t-1$. 
It has been shown that the original NG dynamics and the listener-only version on the complete graph have qualitatively similar results \cite{baronchelliRoleFeedbackBroadcasting2011}. It is easier to derive the iterative function by considering only the state change of listeners, so, we develop our framework for the listener-only version.   

For an uncommitted agent to adopt a single opinion $i$ at step $t$, it must have held the opinion $i$ in its list at step $t-1$ and received opinion $i$ at step $t$. 
By unifying all mixed states that contain opinion $i$ into one variable, $x_{i+}$, such requirements are outlined by Eq.~\myref{eq:x_i}. The first term describes the scenario when a listener already holding the single opinion $i$ receives the signal $i$, and the second term corresponds to the scenario when a listener in the mixed state $x_{i+}$ hears opinion $i$. After the interaction, the listener in both scenarios either remains in the single state $i$ or adapts to it. 
Next, one can establish the recursive relationship of the mixed state containing two opinions, $i$ and $j$, in Eq.~\myref{eq:x_ij}. Specifically, if a listener initially supports opinion $i$ ($j$) and subsequently receives signal $j$ ($i$), it will switch to the mixed state, $ij$. This equation accounts for the scenario where a listener holds one opinion but is influenced by another received opinion through interaction with other agents. 
Similarly, the recursive relationship of the mixed state containing three single opinions is derived in Eq.~\myref{eq:x_ijk}.
Furthermore, one can easily generalize the iteration function of the mixed state containing $n$ single opinions as Eq.~\myref{eq:x_n}, where $\mathcal{S}_n( i_1, i_2, ..., i_n)$ represents all permutations of a set containing $n$ elements.

\begin{equation}
	x_i^{(t)} = x_i^{(t-1)} Q_i^{(t-1)}  + x_{i+}^{(t-1)} Q_i^{(t-1)} \label{eq:x_i} 
\end{equation}

\begin{equation}
	x_{ij}^{(t)} = x_i^{(t-1)} Q_j^{(t-1)} + x_j^{(t-1)} Q_i^{(t-1)} \label{eq:x_ij} 
\end{equation}

\begin{equation}
\begin{split}
	x_{ijk}^{(t)} =& x_{ij}^{(t-1)} Q_k^{(t-1)} + x_{ik}^{(t-1)} Q_j^{(t-1)} + x_{jk}^{(t-1)} Q_{i}^{(t-1)} \\
		      =& x_i^{(t-2)} Q_j^{(t-2)} Q_k^{(t-1)} + x_j^{(t-2)} Q_i^{(t-2)} Q_k^{(t-1)}  \\
		       &+ x_i^{(t-2)} Q_k^{(t-2)} Q_j^{(t-1)}  + x_k^{(t-2)} Q_i^{(t-2)} Q_j^{(t-1)} \\
		      &  + x_j^{(t-2)} Q_k^{(t-2)} Q_i^{(t-1)} + x_k^{(t-2)} Q_j^{(t-2)} Q_i^{(t-1)} \\
		      =& \sum_{(i', j', k') \in \mathcal{S}_3(i, j, k)}^{} x_{i'}^{(t-2)} Q_{j'}^{(t-2)} Q_{k'}^{(t-1)}
\end{split} \label{eq:x_ijk} 
\end{equation}

\begin{equation}
\begin{split}
	x_{i_1i_2...i_n}^{(t)} &= \sum_{(i'_1, i'_2, ..., i'_n) \in \mathcal{S}_n(i_1, i_2, ..., i_n)}^{} x_{i'_1}^{(t-n+1)} \\
			       & \times Q_{i'_2}^{(t-n+1)} Q_{i'_3}^{(t-n)} ... Q_{i'_{n-1}}^{(t-2)} Q_{i'_n}^{(t-1)}. \label{eq:x_n}
\end{split}
\end{equation}

\subsection{The focus on single opinions}
To simplify the computation and focus on the density distribution of single opinions, $x_i$, the need to calculate or record all mixed states is eliminated. Instead, only  $Q_i$ and $x_{i+}$ need to be tracked.  The density evolution of mixed states containing opinion $i$, such as $x_{i\tilde{i}}$,  $x_{i\tilde{i}\tilde{i}}$, $x_{i\tilde{i}\tilde{i}\tilde{i}}$, can be derived using Eq.~\myref{eq:x_n}, where $\tilde{i}$ refers to any single opinion other than opinion $i$.
Therefore, the number of variables is reduced from $2^m-1$ to $m^2$.

By summing up Eq.~\myref{eq:x_ij} over a subset that includes any single opinion $j$ other than $i$, one can obtain $x_{i\tilde{i}} ^ {(t)}$ as Eq.~\myref{eq:x_itildei}, where $\mathcal{M}$ is the set of $m$ single opinions, and $\mathcal{M} \setminus i$ represents the set of all single opinions excluding opinion  $i$.

\begin{equation}
	x_{i\tilde{i}}^{(t)} = x_i^{(t-1)} \sum _{j \in \mathcal{M} \setminus i}^{}Q_j^{(t-1)} + Q_i^{(t-1)} \sum _{j \in \mathcal{M} \setminus i}^{}x_j^{(t-1)} \label{eq:x_itildei}
\end{equation}

\begin{figure*}[h!]
	\centering
	\subfloat[ $P_A=0.02$]{
	\includegraphics[width=0.31\textwidth]{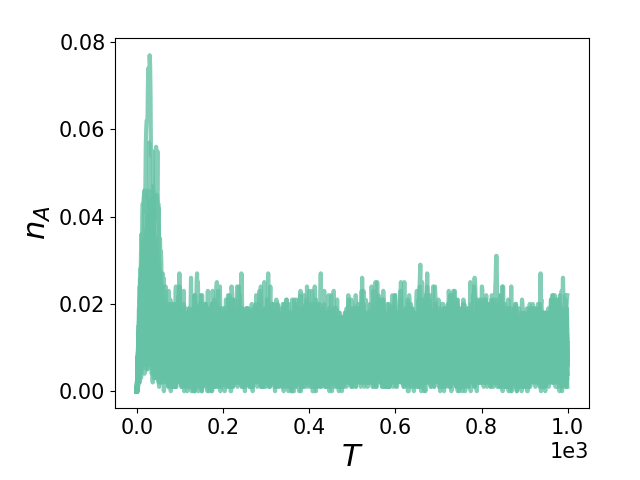}} 
	\subfloat[$P_A=0.03$]{
	\includegraphics[width=0.31\textwidth]{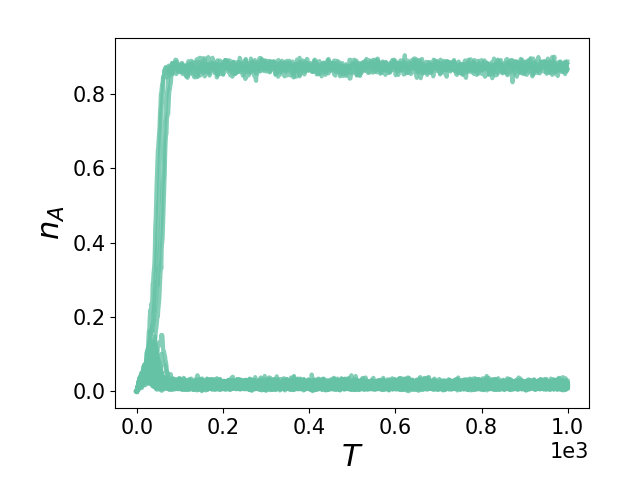}} 
	\subfloat[$P_A=0.04$]{
	\includegraphics[width=0.31\textwidth]{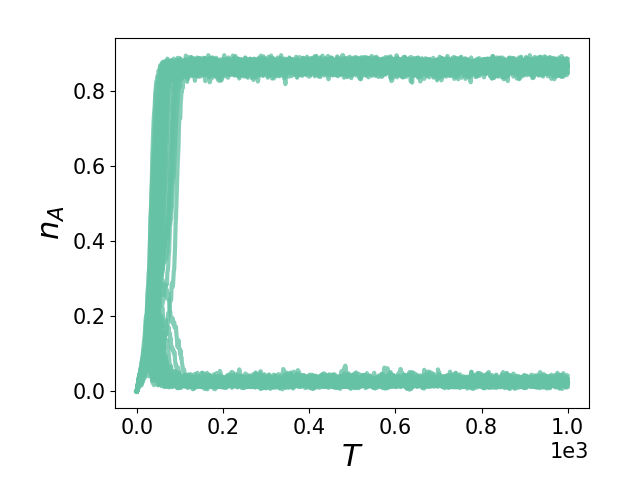}} 
	\caption{\textbf{Time evolution of the fraction of agents supporting opinion $A$ on ER networks.}
 The system comprises $N=1000$ agents, with an average degree of $\langle k \rangle =8$ and an interaction time of $T=1000$. There are $m=5$ single opinions, with a committed fraction of $p_0=0.01$ for each opinion in group $\tilde{A}$, and different values of the fraction committed to opinion $A$ for (a) $P_A = 0.02$, (b) $P_A=0.03$, and (c) $P_A=0.04$. In each panel, there are $L=50$ random realizations, with each line representing one of these realizations.} \label{fig:ER_nA_t}
\end{figure*}

Similarly, one can derive the general formula for the mixed state of length $n+1$ with opinion $i$ and other $n$ distinct opinions,   $x^{(t)}_{i\underbrace{\!\scriptstyle\tilde{i}	\hspace{2pt}...\hspace{2pt}\tilde{i} }_{n}}$,

\iffalse
\begin{equation}
\begin{split}
	x_{i\tilde{i}\tilde{i}}^{(t)} &= x_i^{(t-2)} \left(\sum_{(j, k) \in \mathcal{M} \setminus i}^{} Q_j^{(t-2)} Q_k^{(t-1)}\right)  
				        + Q_i^{(t-2)} \left(\sum_{(j, k) \in \mathcal{M} \setminus i}^{}x_j^{(t-2)} Q_k^{(t-1)}\right) \\
				      &+ Q_i^{(t-1)} \left(\sum_{(j, k) \in \mathcal{M} \setminus i}^{}x_j^{(t-2)} Q_k^{(t-2)}\right)
\end{split}
\end{equation}
\fi

	\begin{equation}
	x^{(t)}_{i\underbrace{\!\scriptstyle\tilde{i}	\hspace{2pt}...\hspace{2pt}\tilde{i} }_{n}}= \sum _{j \in \mathcal{M}  }^{} x_j ^{(t-n)}\sum _{  i \in (j_1,..., j_n) \in \mathcal{M} \setminus j}^{} Q_{j_1}^{(t-n)}... Q_{j_n} ^{(t-1)}
	 \label{eq:x_intildei}
\end{equation}

In Eq.~\myref{eq:x_intildei}, $j_1$, ...,  $j_n$ are $n$ distinct integers, representing $n$ different single opinions. By definition, opinion $i$ must be one of $n$ distinct single opinions $j_1, ..., j_n$.

The primary aim is to monitor the temporal evolution of single opinions, as captured by Eq.~\myref{eq:x_i}. This requires computing the probability of transmitting opinion $i$, $Q_i^{(t)}$, and the density of mixed states,  $x^{(t)}_{i+}$, ($i=1, 2, ..., m$) at each interaction step $t$. According to the interaction rule, only speakers with a single opinion $i$ in their list can communicate opinion $i$. Additionally, for the mixed state, each single opinion in the list has an equal probability of being transmitted. Therefore, $Q_i^{(t)}$ and $x^{(t)}_{i+}$ are expressed as Eqs.~\myref{eq:Q_i} and \myref{eq:x_i+}, respectively.
\begin{equation}
	Q_i^{(t)} = x_i^{(t)} + P_i^{(t)} + \frac{1}{2} x_{i\tilde{i}}^{(t)} +  \frac{1}{3} x_{i\tilde{i}\tilde{i}}^{(t)} +  ... + \frac{1}{m} x^{(t)}_{i\underbrace{\!\scriptstyle\tilde{i}\tilde{i}\tilde{i}	\hspace{2pt}...\hspace{2pt}\tilde{i} }_{n}} \label{eq:Q_i}
\end{equation}

\begin{equation}
\begin{split}
	x_{i+}^{(t)} = x_{i\tilde{i}}^{(t)} +  x_{i\tilde{i}\tilde{i}}^{(t)} +  x_{i\tilde{i}\tilde{i} \tilde{i}}^{(t)} +... + x^{(t)}_{i\underbrace{\!\scriptstyle\tilde{i}\tilde{i}\tilde{i}	\hspace{2pt}...\hspace{2pt}\tilde{i} }_{n}}  \label{eq:x_i+}
\end{split}
\end{equation}

By employing recursive functions \myref{eq:x_i}, \myref{eq:x_intildei}, \myref{eq:Q_i}, and \myref{eq:x_i+}, one can calculate the density evolution of single opinions for any initial condition, offering computational efficiency compared to mean-field differential equations.
Comparing the system evolution obtained by two approaches in Fig.~\ref{fig:compare_recursive_ode}, we find that the results are nearly identical, validating the recursive approach.
One can further simplify the computation if the system's stable state is of primary interest, which means that the probabilities of communicating opinion $i$ at different time steps are the same. Therefore, these probabilities $Q_i^{(t)}$, $Q_i^{(t-1)}$,..., $Q_i^{(t-n)}$ can be represented by one quantity $Q_i^{(s)}$.

\section{The multi-opinion system on random networks}

\begin{figure*}
	\centering
	\subfloat[the fraction of agents supporting $A$]{
	\includegraphics[width=0.45\textwidth]{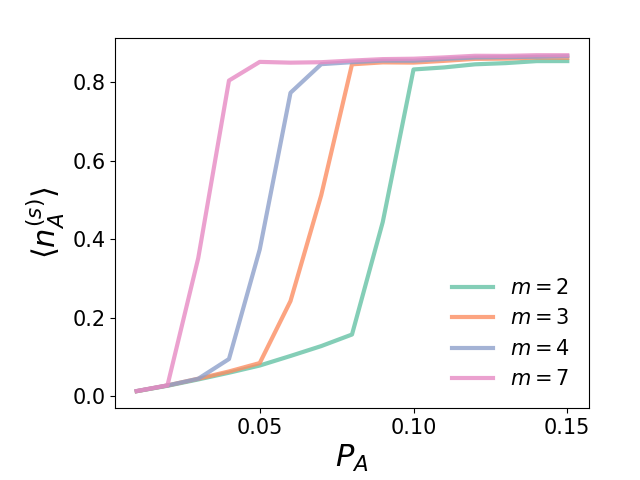}} 
	\subfloat[the ratio of realizations being dominated by $A$]{
	\includegraphics[width=0.45\textwidth]{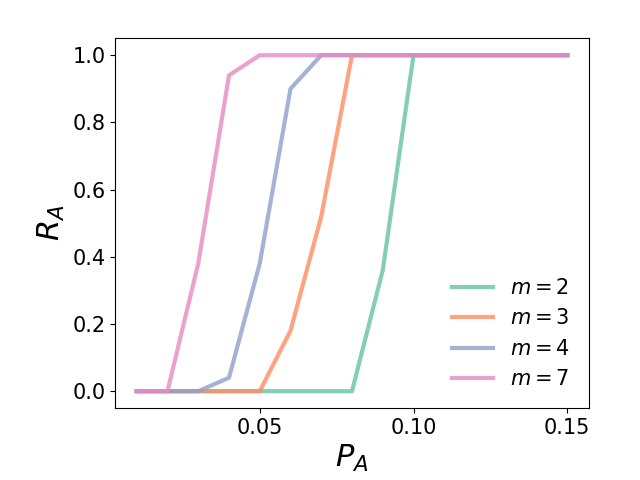}}
	\caption{\textbf{The state of opinion $A$ changes with its committed fraction $P_A$ for different values of $m$ on ER networks.} The system consists of $N=1000$ agents with an average degree of $\langle k \rangle =8$. The fraction of committed agents in group $\tilde{A}$ is $P_{\tilde{A}} = 0.06$. (a) The average fraction of agents supporting opinion $A$ at steady state changes with $P_A$. (b) The ratio of the random realizations that end up with $A$ as the dominant state. Both quantities, $\langle n_A^{(s)} \rangle$ and $R_A$, exhibit a discontinuous transition with $P_A$, and the critical points decrease as the number of opinions increases. This again validates the ``divide and conquer'' phenomenon, as $A$ becomes easier to dominate when the unified group has more divided opinions.} \label{fig:ER_nA_RA_PA}
\end{figure*}

In previous sections, our focus was on understanding opinion dynamics within a complete graph. However, real-world communication often occurs within complex networks. Hence, in this section, we delve into investigating the NG dynamics across diverse network models characterized by real-world features. We aim to identify the tipping point for the dominance transition.
While, in principle, it is possible to develop a heterogeneous (degree-based) mean-field approximation scheme \cite{vespignaniModellingDynamicalProcesses2012,Zhang_PRE2012}, we do not pursue that approach here. Instead, we resort to the agent-based simulation (i.e., using node-based local update rules) to study the density evolution of agents supporting different opinions more precisely. 
On a networked system, agents can be chosen as either speakers or listeners and communicate their opinions with one of their neighbors at each interaction step. Following this exchange, their opinion states are updated according to the NG rule. To set up the agent-based simulation, we employ a system size of $N=1000$, a simulation time of $T=1000$ (defined as the number of pairwise interactions for each agent on average), and conduct $L=50$ random realizations unless specific parameter choices are provided. All committed agents are selected uniformly at random in this study. We acknowledge that the different strategies for allocating the committed minorities on the network can yield varying results \cite{luNamingGameSocial2009}, especially for heterogeneous networks. A systematic exploration of these strategies is beyond the scope of this paper.

We examine a problem similar to what we discussed in the previous sections, with a slight variation: all single opinions now have committed supporters. The opinion with the largest committed fraction is denoted as $A$. For simplicity, when the other $m-1$ opinions share the same fraction of committed fraction, $p_0$, and are initially supported by the same number of uncommitted agents, they can be classified into one group by symmetry, denoted as $\tilde{A}$ with the total committed fraction $P_{\tilde{A}} = (m-1) p_0$. For the finite networked system, either the opinion $A$ or one of the opinions in $\tilde{A}$ would dominate the system in the steady state. 
Our focus lies on determining the critical point, $P_A^{(c)}$, at which opinion $A$ achieves dominance, and understanding how the number of single opinions, $m$, influences this critical point.

\subsection{The impact of random communication topology -- ER networks}
The first model we explore is Erd\H{o}s-R\'enyi (ER) networks \cite{Erdos:1960} because of its wide research interests. 
As agents have different connectivity in random networks and the system size is finite, the evolution and the dominant opinion in the stable state can vary slightly from one realization to another, observed from Fig.~\ref{fig:ER_nA_t}. This variability arises also due to the random selection order of agents as speakers and listeners. These factors introduce randomness in finite systems, resulting in variations in the system's behavior.

To represent the system state, the average fraction $\langle n_i \rangle $ of agents supporting the opinion $i$ is defined in Eq.~\myref{eq:n_i}, where $L$ is the number of realizations. Additionally, we introduce the ratio $R_i$ as the fraction of realizations that end up being dominated by opinion $i$. 
\begin{equation}
	\langle n_i \rangle = \frac{1}{L} \sum _{j=1}^{L} n_i ^{(j)} \label{eq:n_i}
\end{equation}

Fig.~\ref{fig:ER_nA_RA_PA} shows that as the committed fraction $P_A$ increases, there is a critical transition from a low density to the dominant state for the average fraction of agents holding opinion $A$, $\langle n_A^{(s)} \rangle $, as well as for the ratio $R_A$.  
To further investigate the transition on ER networks, we define the critical point on random networks, denoted by $P_A^{(c)}$, as the smallest committed fraction that enables the transition ratio $R_A$ to exceed $\frac{1}{2}$ (Note that our chosen conventional cutoff value $\frac{1}{2}$ does not affect the findings). 
To analyze the relationship between the average degree $\langle k\rangle $ and the critical point $P_A^{(c)}$ on random networks, we examined complete graphs and networks with varying average degree $\langle k\rangle$, as shown in Fig.~\ref{fig:pAc_m_ratio} a. Our results indicate that as the number of single opinions $m$ increases, the critical point $P_A^{(c)}$ decreases, in line with the divide-and-conquer policy. Additionally, we observed that the critical point decreases as the average network degree decreases, suggesting that sparse random communication structures may amplify the impact of committed members on the system, such that opinion $A$ with the largest committed fraction is easier to dominate. This phenomenon has been observed in the two-opinion NG system \cite{Zhang_PRE2012}, and it has been also reported in other social dynamics, including innovation spreading dynamics \cite{montanari_spread_2010, zino_two-layer_2020} and evolutionary games \cite{ellison_learning_1993}.

\begin{figure*}
	\centering
	\subfloat[ER]{
	\includegraphics[width=0.33\textwidth]{{heatmap_ER_N=1000_pAtilde=0.12_pA_critical_m_ratio}}} 
	\subfloat[SW]{
	\includegraphics[width=0.33\textwidth]{{heatmap_SW_N=1000_pAtilde=0.12_pA_critical_m_ratio}}} 
	\subfloat[SF]{
	\includegraphics[width=0.33\textwidth]{{heatmap_BA_N=1000_pAtilde=0.12_pA_critical_m_ratio}}} 
	\caption{\textbf{Heat map of the critical point $P_A^{(c)}$ changes with the number of single opinions and the average degree of networks.} (a) Erd\H{o}s-R\'enyi (ER) networks, (b) small-world (SW) networks, and (c) scale-free (SF) networks. The number of agents is $N=1000$. The total fraction of committed agents in the group $\tilde{A}$ is $P_{\tilde{A}} = 0.06$. The critical point is the smallest committed fraction which enables half of the realizations to stabilize with opinion $A$ as a dominant state. The critical point increases as the average degree increases. It indicates that sparse random communication structures can amplify the impact of committed members, such that opinion $A$ is easier to dominate.}  \label{fig:pAc_m_ratio}
\end{figure*}

\subsection{Critical points in different types of random networks}

Having analyzed how the degree of random networks affects the evolution of stable states in our system, we next look into this evolution for random networks with real-world characteristics. ER random networks are often noted as not reflecting the properties displayed in many real-world networks, like power-law degree distribution and small-world connectivity, and varied nodes clustering. To understand how the stable state might evolve in real-world networks, we extend our analysis to scale-free and small-world networks.

Scale-free networks, like those generated by the Barabasi-Albert model \cite{barabasi1999emergence}, are networks with a power-law degree distribution. While there is debate on how pervasive these properties are in the real world, they are found in many technological and biological networks \cite{broido2019scale} with a common example being the World Wide Web. 
Small-world networks, like those generated by the Watts-Strogatz model \cite{watts1998collective}, have high clustering coefficients and low average path lengths \cite{telesford2011ubiquity}, which are properties seen in many real-world networks like social networks, telecommunications networks, and brain networks. We test these two properties by generating scale-free networks with the Barabasi-Albert model and small-world networks using the Watts-Strogatz model. We utilize the NetworkX library \cite{hagberg2008exploring} to implement these generators.

\begin{comment}
\begin{figure*}
	\centering
	\subfloat[Small-world, $N=1000$]{\includegraphics[width=0.45\textwidth]{figure/SW_N=1000_p=0.06_pA_critical_m_ratio.png}}
	\subfloat[LFR, $N=1000$]{\includegraphics[width=0.45\textwidth]{figure/LFR_N=1000_p=0.06_pA_critical_m_ratio_netseed=40.png}}
	\caption{\textbf{Comparing the critical point $P_A^{(c)}$ with small-world and LFR networks of varying degree.} The number of agents is $N=1000$ in (a) and (b). The total fraction of committed agents in the group $\tilde{A}$ is $P_{\tilde{A}} = 0.06$. The critical point is the smallest committed fraction which enables half of the realizations to stabilize with opinion $A$ as a dominant state. The critical point increases as the average connectivity density increases. } \label{fig:critical_sw_lfr}
\end{figure*}

\end{comment}

In Fig. \ref{fig:pAc_m_ratio} we show the critical points as a function of the number of single opinions $m$ and the average degree of the generated scale-free and small-world networks. For these more complex networks, the general relationship between the number of opinions, the average degree, and the critical point remains the same. For all network types, we see that either the decrease in the average degree of networks or the increase in the number of opinions can lower the critical point.

We also see some interesting behaviors specific to the new network structures. For scale-free networks, the critical point is slightly smaller than that seen in the ER networks. The presence of hub nodes in scale-free networks, where none exist in the other two types, explains this decrease in critical points. The power-law degree distribution gives us several highly connected nodes in the network that can facilitate the quick spreading of a single opinion throughout the network. Many poorly connected nodes need a lot of time to succeed in propagating their opinions to other nodes. Hence, they often end up adopting one of the opinions frequently propagated by the hubs.

For the small-world networks, the critical point is significantly higher across the board. This is due, in part, to the significantly larger clustering present in these networks. Higher clustering allows for single opinions to become entrenched in locally dense portions of the network. Once entrenched, these opinions become harder to unseat, due to the weaker connectivity around the cluster. It is even more difficult, in highly modular networks, for the system to converge to a single dominant opinion \cite{luNamingGameSocial2009}.

\section{Discussions}

In this study, we focus on the competition of the opinion with the largest fraction of committed agents against other opinions with committed agents and the opinion with the majority of uncommitted supporters. We study such competition using the original NG dynamics and its listener-only version. 
While continuous-time mean-field differential equations can accurately describe the opinion evolution for complete graphs in the infinite-size limit, the complexity of systems with multiple opinions grows exponentially, making direct integration of the corresponding differential equations impractical. 

To address this challenge, we introduce two simplified scenarios, $S_1$ and $S_2$, which feature more symmetric setups. These scenarios significantly reduce computational complexity and provide upper and lower bounds for the critical point ($P_A^{(c)}$) of dominance transition in the scenario with an arbitrary distribution of committed agents.
Through comparative analysis of critical transitions across the three scenarios, we highlight the significant influence of the distribution of committed agents within the minority committed group, $\tilde{A}$, in determining $P_A^{(c)}$. Specifically, the number of opinions and the distribution pattern of committed agents within group $\tilde{A}$ can either facilitate or hinder the propagation and eventual dominance of opinion $A$ over uncommitted agents. When opinion $B$ without committed followers is the primary competitor, augmenting the number of committed agents in $\tilde{A}$ can lower $P_A^{(c)}$ by diminishing the support for opinion $B$. Conversely, if agents committed to opinions other than $A$ are the main opponents, increasing their number requires a higher fraction of agents committed to $A$, thereby raising the critical point.

Furthermore, to enhance the accuracy of depicting the NG opinion dynamics and capture critical transitions across various initial conditions in a computationally manageable manner, we develop the discrete-time recursive approach. This method focuses more on the evolution of single opinions by consolidating mixed states with the same opinion into a single variable and introducing the probability of a randomly chosen speaker communicating any single opinion. 
By streamlining computations while preserving the system's dynamics, this framework offers an efficient representation of NG dynamics in a complete graph.

Additionally, to gain insights into opinion evolution within real-world structures, we conducted agent-based simulations to understand system dynamics and capture critical transitions across various finite-sized networks. In our experimental setup, the primary committed group advocates for opinion $A$, while the remaining agents, both committed and uncommitted, are evenly distributed among other minor committed opinions. Our observations reveal a strategy akin to the divide-and-conquer policy, where dividing agents into more minor groups results in a reduced critical fraction of agents committed to $A$ required for system dominance. This phenomenon suggests that segmenting agents facilitate easier domination of the opinion with the largest committed size in the system.

While we presented two frameworks to simplify the multi-opinion NG model, there are some limitations to this work. Firstly, extending the theoretical analysis to networks of various topologies would provide a more comprehensive understanding of opinion dynamics in real-world scenarios. Secondly, we can introduce varied commitments to allow individuals to stick to a single opinion temporarily while maintaining their long-term flexibility, particularly relevant for moderately committed agents. Thirdly, we would also like to extend the original Naming Game model from pairwise interactions to group interactions, allowing for the consideration of discussions within groups of friends, which is common in real-life situations.

\section*{Acknowledgements}
B.K.S. was partially supported by DARPA-INCAS under Agreement No. HR001121C0165 and by the NSF Grant No. BSE-2214216

\bibliographystyle{IEEEtran}

\bibliography{manuscript_v1}% Produces the bibliography via BibTeX.

\begin{IEEEbiography}[{\includegraphics[width=1in,height=1.25in,clip,keepaspectratio]{CM.jpg}}]%
{Cheng Ma
	\normalfont received the Ph.D. degree in Physics from Rensselaer Polytechnic Institute (RPI), Troy, NY in 2024, and he is a member of the Network Science and Technology (NEST) Center at RPI. His current research focuses on network science, computational social systems, and theoretical biology.}
\end{IEEEbiography}

\begin{IEEEbiography}[{\includegraphics[width=1in,height=1.25in,clip,keepaspectratio]{BC}}]%
{Brendan Cross
	\normalfont is a PhD student in the Computer Science department at Rensselaer Polytechnic Institute (RPI), Troy, NY, and a member of the Network Science and Technology (NEST) Center at RPI. His research focuses on computational social systems and community detection.}
\end{IEEEbiography}

\begin{IEEEbiography}[{\includegraphics[width=1in,height=1.25in,clip,keepaspectratio]{GK}}]%
	{Gyorgy Korniss
 \normalfont 
received the diploma degree in physics from Eötvös University, Budapest, Hungary, in 1993, and the Ph.D. degree in physics from Virginia Tech, Blacksburg, VA, USA, in 1997. 
\\
\indent He was a Postdoctoral Research Associate with the Supercomputer Computations Research Institute, Florida State University, Tallahassee, FL, USA, from 1997 to 2000. He has been with the Department of Physics, Rensselaer Polytechnic Institute, Troy, NY, USA, since 2000, where he has been a Full Professor since 2012. His research background is statistical physics, computational physics, and complex systems. His research focuses on opinion dynamics and influencing in social networks, transport, flow, and cascading failures in complex networks, and synchronization, coordination, and extreme events in coupled stochastic systems. He served on the Editorial Boards of {\it Scientific Reports}, from 2014 to 2023, {\it Fluctuation and Noise Letters}, from 2007 to 2022, and {\it Computational Social Networks},
from 2014 to 2016. 
 }
\end{IEEEbiography}

\begin{IEEEbiography}[{\includegraphics[width=1in,height=1.25in,clip,keepaspectratio]{BKS}}]%
	{Boleslaw K. Szymanski
\normalfont
(Life Fellow, IEEE) received the M.Eng. degree in electronics from Warsaw Polytechnic Institute, Warsaw, Poland, in 1973, and the Ph.D. degree in computer science from the Informatics Institute, National Academy of Sciences, Warsaw, Poland, in 1976. 
\\
\indent He is the Claire and a Roland Schmitt Distinguished Professor in computer science, a Professor in physics, and the Founding Director of the Network Science and Technology (NEST) Center, Rensselaer Polytechnic Institute (RPI), Troy, NY, USA. His research focuses on computational social systems, computer networks, and distributed computing. He is a IEEE Fellow (1999) and a AAAS Fellow (2024). 
\\
\indent Prof. Szymanski received Wilkes Medal from the British Computer Society, the Service Award from Network Science Society and since 2009 is a Foreign Member of the National Academy of Science in Poland. 
}
\end{IEEEbiography}

\end{document}